\documentclass[sigconf]{acmart}
\usepackage{url}
\usepackage{multirow}
\usepackage{balance}
\usepackage{rotating}
\usepackage{booktabs}
\usepackage{hyphsubst}
\usepackage{graphicx}
\usepackage{amsmath}
\graphicspath{{images/}}
\usepackage{times}
\usepackage{hhline}
\usepackage[caption = false]{subfig}

\usepackage{colortbl}

\usepackage{url}
\usepackage{rotating}
\usepackage{amsmath}
\usepackage{cancel}
\usepackage{amssymb}
\usepackage{enumitem}
\usepackage{pifont}

\usepackage{multirow}
\usepackage{array}
\usepackage{placeins}

\usepackage{listings}

\usepackage{color}

\newcommand\YAMLcolonstyle{\color{orange}\mdseries}
\newcommand\YAMLkeystyle{\color{black}\bfseries}
\newcommand\YAMLvaluestyle{\color{blue}\mdseries}

\lstdefinestyle{json}
{
    string=[s]{"}{"},
    stringstyle=\color{black},
    comment=[l]{:},
    commentstyle=\color{blue},
}

\makeatletter

\newcommand\language@yaml{yaml}

\expandafter\expandafter\expandafter\lstdefinelanguage
\expandafter{\language@yaml}
{
  keywords={true,false,null,y,n},
  keywordstyle=\color{blue},
  basicstyle=\YAMLkeystyle,                                 
  sensitive=false,
  comment=[l]{\#},
  morecomment=[s]{/*}{*/},
  commentstyle=\color{red}\ttfamily,
  stringstyle=\YAMLvaluestyle\ttfamily,
  moredelim=[l][\color{orange}]{\&},
  moredelim=[l][\color{magenta}]{*},
  moredelim=**[il][\YAMLcolonstyle{:}\YAMLvaluestyle]{:},   
  morestring=[b]',
  morestring=[b]",
  literate =    {---}{{\ProcessThreeDashes}}3
                {>}{{\textcolor{orange}\textgreater}}1     
                {|}{{\textcolor{orange}\textbar}}1 
                {\ -\ }{{\mdseries\ -\ }}3,
}

\lst@AddToHook{EveryLine}{\ifx\lst@language\language@yaml\YAMLkeystyle\fi}
\makeatother

\newcommand\ProcessThreeDashes{\llap{\color{cyan}\mdseries-{-}-}}

\definecolor{pblue}{rgb}{0.13,0.13,1}
\definecolor{pgreen}{rgb}{0,0.5,0}
\definecolor{pred}{rgb}{0.9,0,0}
\definecolor{pgrey}{rgb}{0.46,0.45,0.48}

\usepackage{rotating}
\usepackage{booktabs}
\usepackage{rotating}
\usepackage{hhline}
\usepackage{hyphsubst}
\usepackage{soul}

\newlength{\Oldarrayrulewidth}

\copyrightyear{2018}
\acmYear{2018}
\setcopyright{rightsretained}
\acmConference{IFUP Workshop @ ACM WSDM '18}{}{February 6, 2018, Los Angeles, USA}
\acmDOI{N/A}
\acmISBN{}
\acmPrice{}
\begin{document}

\title{Beyond Accuracy Optimization: On the Value of Item Embeddings for Student Job Recommendations}

\author{Emanuel Lacic}
\affiliation{%
  \institution{Know-Center Graz}
  \city{Graz} 
  \state{Austria} 
}
\email{elacic@know-center.at}

\author{Dominik Kowald}
\affiliation{%
  \institution{Know-Center Graz}
  \city{Graz} 
  \state{Austria} 
}
\email{dkowald@know-center.at}

\author{Markus Reiter-Haas}
\affiliation{%
  \institution{Moshbit GmbH}
  \city{Graz} 
  \state{Austria} 
}
\email{markus.reiter-haas@studo.co}

\author{Valentin Slawicek}
\affiliation{%
  \institution{Moshbit GmbH}
  \city{Graz} 
  \state{Austria} 
}
\email{valentin.slawicek@studo.co}

\author{Elisabeth Lex}
\affiliation{%
  \institution{Graz University of Technology}
  \city{Graz} 
  \state{Austria} 
  }
\email{elisabeth.lex@tugraz.at}

\begin{abstract}
In this work, we address the problem of recommending jobs to university students. For this, we explore the utilization of neural item embeddings for the task of content-based recommendation, and we propose to integrate the factors of frequency and recency of interactions with job postings to combine these item embeddings. We evaluate our job recommendation system on a dataset of the Austrian student job portal Studo using prediction accuracy, diversity and an adapted novelty metric. This paper demonstrates that utilizing frequency and recency of interactions with job postings for combining item embeddings results in a robust model with respect to accuracy and diversity, which also provides the best adapted novelty results.
\end{abstract}


\maketitle

\begin{figure}[t!]
\centering
\subfloat{
  \centering
  \includegraphics[width=.40\columnwidth]{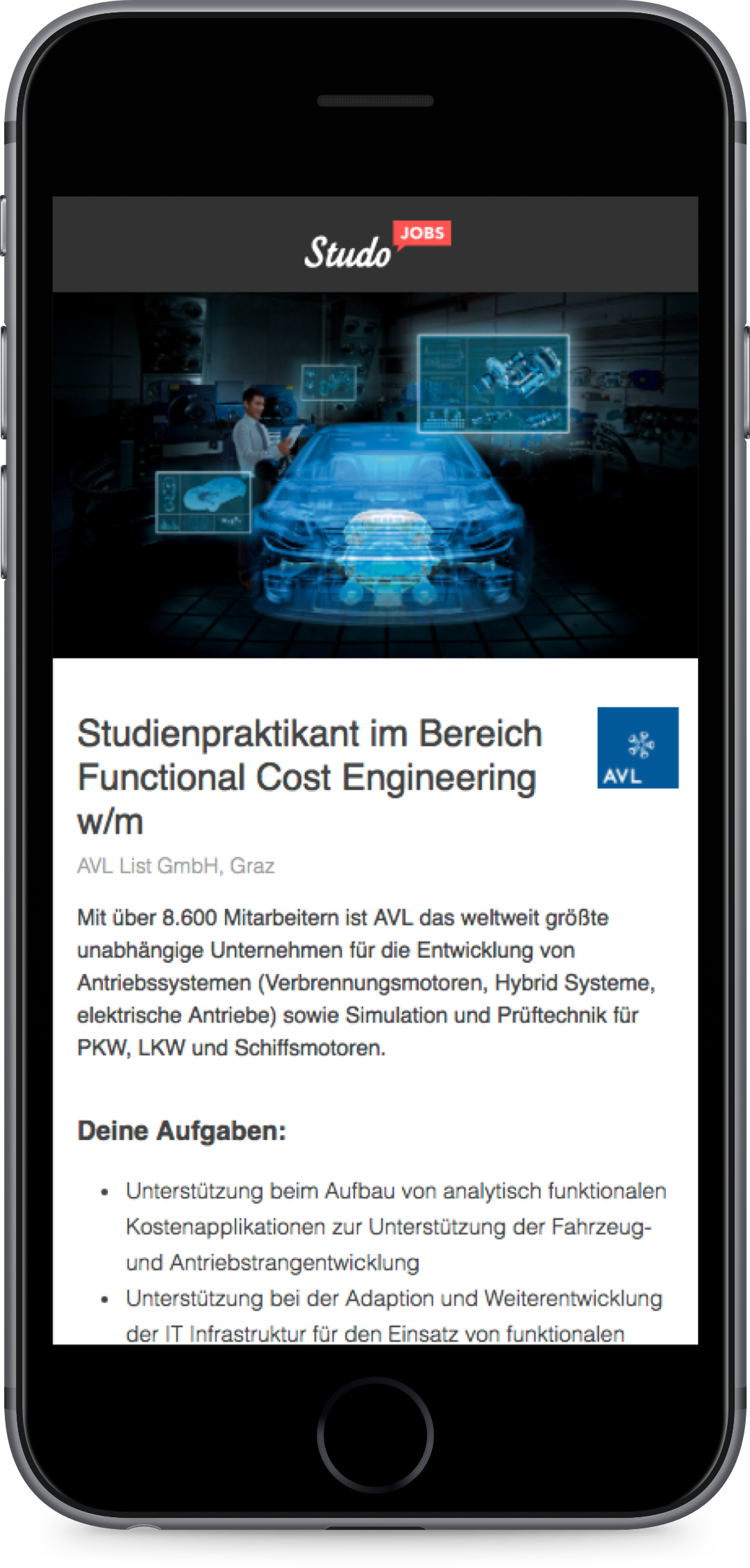}
  \label{fig:detailview}
}%
\subfloat{
  \centering
  \includegraphics[width=.40\columnwidth]{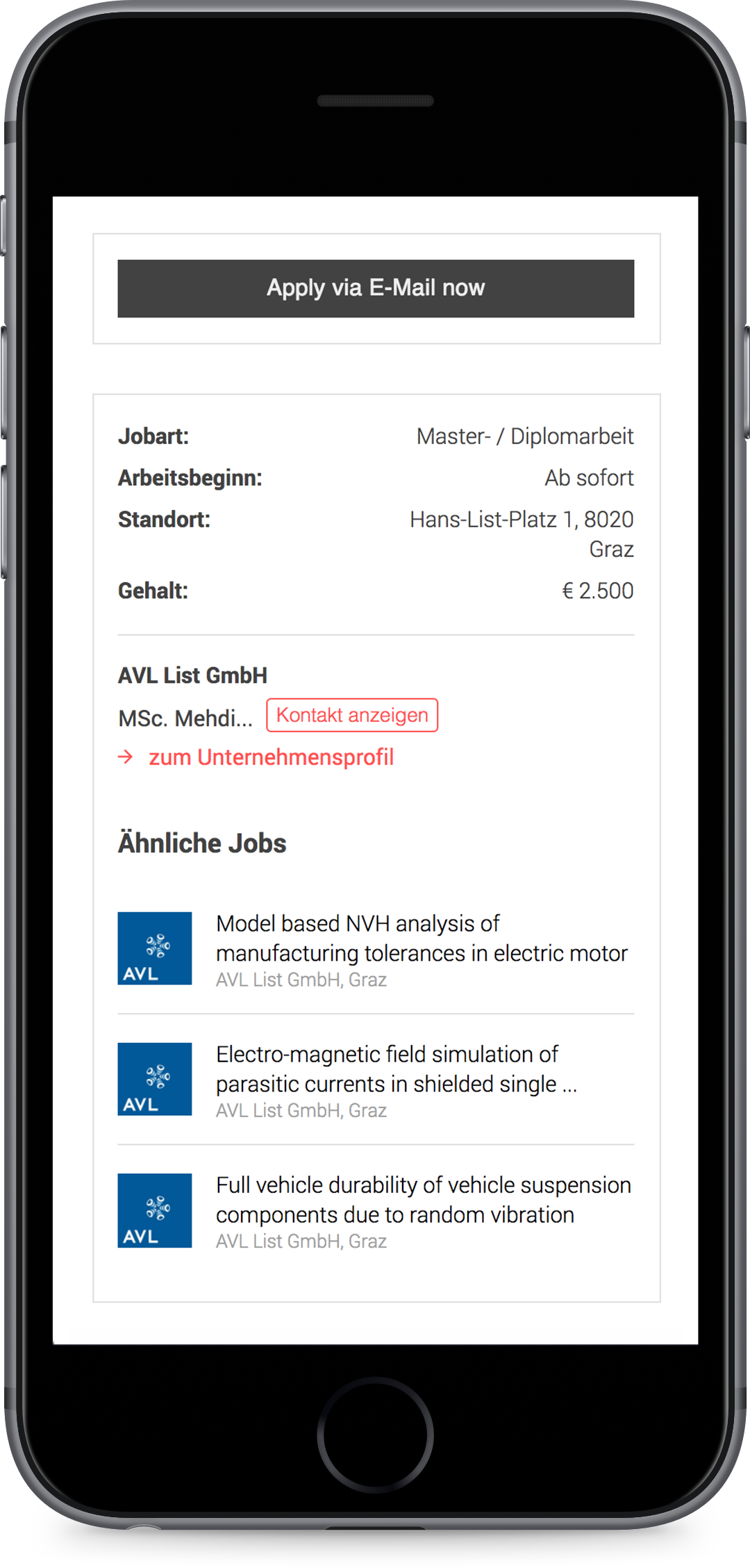}
  \label{fig:metainfo}
}
\caption{Screenshot of the Studo mobile application, which shows (a) the detail view of a job posting and, (b) a recommendation with low diversity.}
\label{fig:studo}
\vspace{-0.5cm}
\end{figure}

\section{Introduction}
\label{sec:introduction}

Finding jobs is not a trivial task for recently graduated university students since they usually only have a little or no relevant work experience at all. Nowadays, this is a real issue as having a university degree does not automatically guarantee students to get their desired job after graduation. For example, a recent study \cite{jones2014college} reports that in 2013, $23.5\%$ of employed U.S. graduates are not only underemployed but also work in positions with a below-than-average salary. We recently started to tackle the same problem for Austrian university students within the Studo mobile application\footnote{\url{https://studo.co/}}. Studo provides guidance and support for about one third of Austrian students 
by offering services such as finding the right job\footnote{\url{https://jobs.studo.co/}} in order to gather relevant work experience before they graduate (see Figure \ref{fig:detailview}).


In Studo, we currently use a content-based job recommendation system to suggest similar jobs to the currently viewed one. Due to limited display space in mobile applications, we focus on a low number of job recommendations (e.g., 3 or 6) that are shown to the user. As such, we frequently encounter the problem of a poor recommendation diversity. This is also demonstrated in Figure \ref{fig:metainfo}, in which companies often reuse a large portion of their job description text for other open positions, which leads to content-based recommendations that the users may perceive as unhelpful. Apart from recommendation diversity, novelty is also an important metric for personalized job recommendations since applying to popular jobs may decrease the student's satisfaction due to high competition and less chance of getting hired (see e.g., \cite{Kenthapadi2017}).

\vspace{2mm}
\noindent 
\textbf{The present work.}
In this work, we apply the  Doc2Vec  algorithm  \cite{le2014distributed} to obtain fixed-length job vectors to improve recommendation diversity and novelty. That is, we learn neural network-based item embeddings 
in an unsupervised manner from the job description. This type of item embeddings enables to represent the job description as a dense vector that takes into account its semantic and syntactic structure. 

In order to construct these vectors, we utilize three approaches. Firstly, we use the vector representation of the job posting the user has most recently interacted with. Secondly, we synthetically create a vector representations of all jobs the user has interacted with in the past by averaging the column values of these vectors. 
Thirdly, we use the Base-Level Learning (BLL) equation from human memory theory \cite{anderson2004integrated} to integrate the factors of how frequently and recently the user has interacted with job postings in the past. 
We evaluate these approaches using a dataset gathered from the job portal used by Austrian students within the Studo mobile application.

\vspace{2mm}
\noindent 
\textbf{Contributions.} 
Our contributions are three-fold: (i) we explore the impact of item embeddings on not only accuracy but also novelty and diversity for recommending jobs to students, (ii) we utilize the BLL equation from human memory theory to model the frequency and recency of job interactions, and (iii) we propose an adapted novelty measure to evaluate job recommendations.



In summary, this study may help researchers to obtain an overview of how to combine learned item embeddings in order to optimize diversity and novelty in addition to accuracy. Our results show that utilizing the BLL equation can improve recommendation performance. Moreover, combining the BLL-induced item embeddings with user-based Collaborative Filtering produces a robust model that gives the best results with respect to our adapted novelty metric.



\section{Related Work}

Research on job recommmendations has mostly focused on improving the accuracy by applying various Collaborative and Content-Based Filtering approaches or their hybrid combinations \cite{al2012survey,zhang2016ensemble}. \cite{Liu2017} inverted the problem and searched for the right users for a given job and then recommended this job to the users. Classic machine learning-based methods have also been used for job recommendations. For example, the work of \cite{Paparrizos2011} recommended jobs based on inference from the job transition patterns observed in the past. This, however, is not directly applicable for student job recommendations as students often have no job experience at all. 

Models that train item embeddings, such as Word2Vec \cite{mikolov2013distributed}, have been shown to provide good results in many natural language processing tasks. As such, research on improving the recommendation performance by employing word embedding vectors is gaining more momentum. 
As an example, the authors of \cite{musto2016learning} used the learned word embeddings trained on Wikipedia for content-based recommender systems. Additionally, in \cite{huang2016exploiting}, word embeddings were employed to represent job postings. In order to achieve a high accuracy, a large number of embedding dimensions were required. 

The authors of \cite{yuan2016solving} utilized item embeddings on top of Collaborative Filtering to tackle the cold-start problem for job recommendations.
In addition to that, in the RecSys Challenge of 2016, \cite{Liu2016} suggested that time-dependent models can enhance the job recommendation performance, which further motivates the applicability of our BLL-based approach.

\section{Approach}

Over the last four years, there has been a number of notable publications in the area of applying deep learning for recommender systems. Especially methods for finding good embeddings (also known as latent representations) for items have become popular. Most models use some variation of Word2Vec \cite{mikolov2013distributed}. In our work, we utilize the DBOW approach of \cite{le2014distributed} (also known as Doc2Vec or Paragraph Vector), which works in the same way as skip-gram, except that the input is replaced by a token representing the job posting.
Thus, we first train our model using the description text of job postings and afterwards extract the corresponding vector representations. 

To train a Doc2Vec model,
we need to set several parameters beforehand. These are the window size (i.e., $5$, $10$, $15$ or $20$), number of negative samples (i.e., $0$, $5$, $10$, $15$) and number of epochs to train (i.e., from $1$ to $5$). 
Based on multiple parameter combination experiments, we trained the final item representations using the job description text with a window size of $20$, a learning rate of $0.025$, $10$ negative samples and $1$ training epoch.

\vspace{2mm}
\noindent 
\textbf{Recommendation strategy.} 
In order to use the extracted vector representations (i.e., embeddings) in a content-based manner, we need to construct a reference vector for which we can find top-$k$ similar vectors (i.e., job postings) using the Cosine similarity. To obtain this reference vector, we initially define two different strategies:

\begin{itemize}
\item[LAST] We use the vector representation of the job posting the user has most recently interacted with. This strategy simulates the current use-case in Studo of recommending similar job postings to the currently viewed one. 
\item[AVG] We use the user's whole job interaction history and synthetically create a reference vector. This is done by averaging the columns of all the vectors that are assigned to the job postings contained in the history.
\end{itemize}

\vspace{2mm}
\noindent 
\textbf{Integrating frequency and recency of job interactions.}
Research on how humans are accessing information in their long-term memory has shown that past interaction frequency and recency are crucial factors \cite{anderson2004integrated}. In this respect, the Base-Level Learning (BLL) equation, defined by the cognitive architecture ACT-R, integrates these two factors in order to model the information access in human memory. Our previous research has shown that the BLL equation can be utilized for developing social tag and hashtag recommender systems \cite{kowald2017temporal}.

One issue of our proposed LAST and AVG approaches is that they either focus on the factors of interaction recency (i.e., LAST) or interaction frequency (i.e., AVG) and thus, do not account for both factors simultaneously. Thus, we use the BLL equation to combine the factors of how frequently and recently the user has interacted with job postings in the past:
\vspace{-0.15cm}
\begin{equation}
	BLL_{j} = \ln(\sum\limits_{i = 1}\limits^{n}{(TS_{ref} - TS_{j,i})^{-d})}
\end{equation}

Here, $BLL_{j}$ is the BLL value for a given user $u$ and a given job $j$, and $n$ states the number of times $u$ has interacted with $j$ in the past. Moreover, $TS_{j,i}$ is the timestamp (in seconds) when $u$ has interacted with $j$ for the $i$th time and $TS_{ref}$ is a reference timestamp such as the time when the job recommendations are requested. $d$ is a parameter to set the time-dependent decay of item exposure in human memory and, according to \cite{anderson2004integrated}, we set it to its default value of $0.5$.

These BLL values are first normalized using a softmax function and then multiplied with the individual job vectors from the user's history. 
This way, we form a weighted sum of job postings based on how frequently and recently the user has interacted with them.




\section{Methodology}
To evaluate the performance of our approach, we followed an offline evaluation methodology. 

\vspace{2mm}
\noindent 
\textbf{Dataset.}
We extracted our data from Studo and split it in two different sets (training and test set) using a method similar to the one described in \cite{lacic2017tailoring}. Thus, for each user with at least $11$ items in the history, we withheld $10$ viewed job postings from the training set and added them to the test set to be predicted. This results in $3,011$ users who get recommendations from $2,345$ job postings. The interaction data contains $140,411$ job views, which results in a sparsity of $98.01\%$. 



\vspace{2mm}
\noindent 
\textbf{Experimental setup.} We utilize three well-known baselines from the literature:

\begin{itemize}
\item[MP] The Most Popular approach recommends for any user the same set of items, which are weighted and ranked by the frequency of job views. 
\item[CBF] Content-Based Filtering analyses item metadata to identify other items that could be of interest for a specific user. This is  done using TF-IDF on the description text of the job the user has most recently interacted with.
\item[CF] User-Based Collaborative Filtering is defined as recommending jobs to a target user that have been previously viewed by similar users (i.e., the neighbors). To find the $k$ nearest neighbors, we utilize the Cosine similarity metric.
\end{itemize}

We did not evaluate Matrix Factorization approaches as our focus lies on calculating 
recommendations in real-time. That is, we do not employ an offline model update strategy that could last hours or days and potentially miss the user’s real-time demand. The neural model, which is used to extract item embeddings does not present an issue here as this operation is performed as soon as a new job postings comes to the system. This means that the job embedding can be used for recommendation purposes as soon as the job is posted.

\def\arraystretch{1.3}
\begin{table*}[ht]
\setlength{\tabcolsep}{10pt}

\centering
\scalebox{0.92}{
\begin{tabular}{l|l|l||c|c|c|c||c|c|c|c}

\multicolumn{3}{c||}{Approach} & \multicolumn{4}{c||}{k = 3}                                                              & \multicolumn{4}{c}{k = 6}                                                             \\  \cline{4-11}
\multicolumn{3}{c||}{}         
& \multicolumn{1}{c|}{nDCG} & \multicolumn{1}{c|}{Novelty} & \multicolumn{1}{c|}{Diversity} & \multicolumn{1}{c||}{Novelty$^*$} 
& \multicolumn{1}{c|}{nDCG} & \multicolumn{1}{c|}{Novelty} & \multicolumn{1}{c|}{Diversity} & \multicolumn{1}{c}{Novelty$^*$}  \\ \hline \hline

\multicolumn{3}{c||}{MP}         
&  \cellcolor{blue!55} $.0395$  &  \cellcolor{cyan!5} $.1649$    & \cellcolor{teal!55} $.7261$  &       \cellcolor{green!5}$.5849$             
&  \cellcolor{blue!55} $.0722$  &  \cellcolor{cyan!5} $.1857$    & \cellcolor{teal!50} $.7156$  &       \cellcolor{green!5}$.6057$             
\\ \hline
\multicolumn{3}{c||}{CBF}        
&  \cellcolor{blue!20} $.0122$  & \cellcolor{cyan!55} $.7676$    & \cellcolor{teal!5} $.4536$   &       \cellcolor{green!25}$.8124$             
&  \cellcolor{blue!20}  $.0156$ & \cellcolor{cyan!65} $.7835$    & \cellcolor{teal!5} $.4854$   &       \cellcolor{green!15}$.7965$             
\\ \hline
\multicolumn{3}{c||}{CF}         
& \cellcolor{blue!65} $.0889$ &  \cellcolor{cyan!10} $.3518$ &  \cellcolor{teal!40} $.6736$     &       \cellcolor{green!15}$.7718$             
& \cellcolor{blue!65} $.1292$ &  \cellcolor{cyan!10} $.3660$ &  \cellcolor{teal!30} $.6814$     &       \cellcolor{green!10}$.7860$             
\\ \hline \hline

\multirow{9}{5pt}{\centering{\begin{sideways}\centering{Doc2Vec}\end{sideways}}}

& \multirow{3}{5pt}{\centering{\begin{sideways}\centering{LAST}\end{sideways}}}

& \multicolumn{1}{c||}{d=100}        
&  \cellcolor{blue!40}  $.0170$    &  \cellcolor{cyan!35} $.7469$     &  \cellcolor{teal!10} $.4845$    &       \cellcolor{green!45}$.8331$             
&  \cellcolor{blue!40}  $.0217$    &  \cellcolor{cyan!45} $.7639$     &  \cellcolor{teal!10} $.5486$    &       \cellcolor{green!35}$.8161$             
\\ \cline{3-11}
&& \multicolumn{1}{c||}{d=200}        
&  \cellcolor{blue!50} $.0182$    &  \cellcolor{cyan!30} $.7389$     &   \cellcolor{teal!15}  $.5091$   &       \cellcolor{green!50}$.8411$             
&  \cellcolor{blue!45} $.0219$    &  \cellcolor{cyan!35} $.7588$     &   \cellcolor{teal!15}  $.5854$   &       \cellcolor{green!45}$.8212$             
\\ \cline{3-11}
&& \multicolumn{1}{c||}{d=300}        
&  \cellcolor{blue!45} $.0177$    &  \cellcolor{cyan!25}  $.7352$    &  \cellcolor{teal!20}  $.5163$    &       \cellcolor{green!55}$.8448$             
&  \cellcolor{blue!50} $.0220$    &  \cellcolor{cyan!40}  $.7594$    &  \cellcolor{teal!20}  $.5953$    &       \cellcolor{green!40}$.8206$             
\\ \cline{2-11}

& \multirow{3}{5pt}{\centering{\begin{sideways}\centering{AVG}\end{sideways}}}  & d=100

& \cellcolor{blue!15} $.0107$  &  \cellcolor{cyan!45} $.7525$     &  \cellcolor{teal!50}  $.6929$       &       \cellcolor{green!35}$.8275$             
& \cellcolor{blue!15}  $.0154$ &  \cellcolor{cyan!50} $.7656$     & \cellcolor{teal!55} $.7239$         &       \cellcolor{green!30}$.8144$             
\\ \cline{3-11}
&& d=200                        
& \cellcolor{blue!10} $.0099$ & \cellcolor{cyan!60}  $.7870$ &  \cellcolor{teal!65}  $.7455$            &       \cellcolor{green!20}$.7930$             
& \cellcolor{blue!10} $.0135$ & \cellcolor{cyan!60}  $.7750$ &  \cellcolor{teal!65}  $.7830$            &       \cellcolor{green!25}$.8050$             
\\ \cline{3-11}
&& d=300                        
&  \cellcolor{blue!5} $.0091$ &  \cellcolor{cyan!65} $.8085$     &  \cellcolor{teal!60} $.7439$         &       \cellcolor{green!10}$.7715$             
&  \cellcolor{blue!5} $.0133$ & \cellcolor{cyan!55}  $.7797$     &  \cellcolor{teal!60}$.7796$          &       \cellcolor{green!20}$.8003$             
\\ \cline{2-11}

& \multirow{3}{5pt}{\centering{\begin{sideways}\centering{BLL}\end{sideways}}}

& d=100                        
& \cellcolor{blue!35} $.0156$ & \cellcolor{cyan!20} $.7300$   & \cellcolor{teal!25}  $.5974$            &       \cellcolor{green!60}$.8500$             
& \cellcolor{blue!35} $.0198$ & \cellcolor{cyan!20} $.7478$   & \cellcolor{teal!25}  $.6486$            &       \cellcolor{green!60}$.8322$             
\\ \cline{3-11}
&& d=200                        
& \cellcolor{blue!30} $.0146$ & \cellcolor{cyan!40} $.7516$   &  \cellcolor{teal!35}$.6408$             &       \cellcolor{green!40}$.8284$             
& \cellcolor{blue!30} $.0188$ & \cellcolor{cyan!25} $.7525$   & \cellcolor{teal!40} $.7006$             &       \cellcolor{green!55}$.8275$             
\\ \cline{3-11}
&& d=300                        
& \cellcolor{blue!25} $.0144$ &  \cellcolor{cyan!50} $.7609$ & \cellcolor{teal!30}$.6388$               &       \cellcolor{green!30}$.8191$             
& \cellcolor{blue!25} $.0186$ &  \cellcolor{cyan!30} $.7578$ & \cellcolor{teal!45} $.7015$              &       \cellcolor{green!50}$.8222$             
\\ \hline \hline

\multicolumn{3}{c||}{Mixed Hybrid}                       
& \cellcolor{blue!60} $.0721$    & \cellcolor{cyan!15} $.4531$ &  \cellcolor{teal!45} $.6820$           &       \cellcolor{green!65}$.8731$             
& \cellcolor{blue!60} $.0900$    & \cellcolor{cyan!15} $.5378$ &  \cellcolor{teal!35} $.6890$           &       \cellcolor{green!65}$.9578$             
\end{tabular}
}
\caption{Evaluation results which show that a recommendation approach that uses the BLL equation provides a good balance between accuracy and diversity while achieving the best performance with respect to the target novelty (i.e., the Novelty$^*$  measure).}
 \label{table:res}
 \vspace{-0.75cm}
\end{table*}

To quantify the performance of each of our recommender approaches, we use a rich set of well-established metrics in recommender systems. Furthermore, we propose to evaluate job recommendations with respect to an adapted novelty measure.

\vspace{2mm}
\noindent 
\textbf{Normalized Discounted Cumulative Gain (nDCG).} nDCG is a ranking-dependent metric that not only measures how many items can be correctly predicted but also takes the position of the items in the recommended list into account \cite{ParraSahebi}. It is calculated by dividing the DCG of the user's recommendations with the ideal DCG value, which is the highest possible DCG value that can be achieved if all the relevant items would be recommended in the correct order.


\vspace{2mm}
\noindent 
\textbf{Diversity.} As defined in \cite{SmythMcClave01}, recommendation diversity can be calculated as the average dissimilarity of all pairs of items in the list of recommended items. Given a distance function $d(i,j)$ that is the dissimilarity between two items $i$ and $j$, $D@k$ is given as the average dissimilarity of all pairs of items in the list of recommended items \cite{SmythMcClave01}. In our experiments, we use Cosine similarity to measure the dissimilarity of two job postings using the textual content from their descriptions. 
  

\noindent 
\textbf{Novelty.} Recommendation novelty can be seen as the ability of a recommender to introduce users to items that have not been experienced before. A recommendation that is accurate but not novel will include items that the user enjoys, but already knows of. Optimizing on novelty has been shown to have a positive, trust-building impact on user satisfaction \cite{pu2011user}. In our experiments, we measure novelty with a normalized metric previously introduced by \cite{zhou2010solving}:

 \begin{align}
		Novelty@k = 1 - \frac{ 1 }{ |U| } \sum\limits_{ u \in U }{\frac{ 1 }{ k } \sum\limits_{ i \in k }{ \frac{ \log_2 (pop_i + 1) }{ \log_2 (pop_{MAX} + 1) } } }
\end{align}

This way we quantify the average information content of job recommendations, where higher values mean that more globally ``unexplored'' items are being recommended. If the likelihood that a user has experienced an item is proportional to its global popularity, this can serve an approximation of true novelty.

\vspace{2mm}
\noindent 
\textbf{Novelty$^*$.} As already mentioned, we aim to optimize the diversity and novelty of recommendations in order to improve long-term user satisfaction.
One issue, however, is the over-optimization of the novelty metric as this basically leads to recommending only items from the far end of the long-tail. As such, we investigated the difference in novelty on job postings that are being viewed by the users to the ones the students have actually applied to. We found that the novelty value of the applied jobs is on average $0.58$. Thus, we propose to evaluate job recommendation novelty with respect to an adapted novelty measure that considers the desired outcome in the live system. The goal is to minimize the distance between the calculated recommendation novelty and the average novelty of job applications $N_{A}$, which is given by:

\begin{align}
Novelty^*@k = 1 - |N_{A} - N@k|
\end{align}
Here, $N_{A}=0.58$ and $N@k$ is the actual calculated novelty.






\section{Results}
\label{sec:res}
In Studo, we show job recommendations via the mobile app, which has limited screen space. As such, in Table \ref{table:res}, we report the evaluation results for $k=3$ and $k=6$ recommended jobs (although the recommendation performance for $k=10$ follows a similar trend).

We found that both MP and CF perform well based on accuracy and diversity but suffer from poor novelty which is far from our target novelty $ N_{A} $ (i.e., $0.58$).
CBF had opposing results to these, thus suffering from poor accuracy and a rather low diversity. This leads to the previously mentioned situation, where a user is confronted with very similar job recommendations, which can be perceived as not very helpful. While CBF has a very high novelty score (e.g., the highest one for $k=6$), it even overshoots the target novelty $N_{A}$ that we want to reach. 

\vspace{2mm}
\noindent 
\textbf{Impact of embeddings for job recommendation.}
By utilizing item embeddings the same way as CBF (i.e., using the LAST strategy), the accuracy and diversity are slightly improved. The novelty on the other hand is lowered, which is actually desirable as we want to minimize the distance to the target novelty $ N_{A} $. While utilizing the AVG strategy we trade the accuracy for the highest diversity results. Increasing the vector dimensions (i.e., 200 and 300), however, leads to higher novelty, which corresponds to a bad performance of the Novelty$^*$ measure.
These results suggest that a trade-off between both approaches is needed to reach a better balance between accuracy, diversity and novelty.

\vspace{2mm}
\noindent 
\textbf{Effect of the BLL equation.}
By utilizing the BLL equation when constructing item embeddings, we account for the factors of how frequently and recently the user has interacted with job postings in the past.
We can see that the accuracy results are only slightly lower than the ones of the LAST approach. The same can be said about the diversity when compared to the AVG approach. The BLL approach still outperforms CBF regarding both metrics. With respect to the target novelty  $N_{A}$, utilizing the BLL equation results in achieving the best Novelty$^*$ performance and this even with the smallest number of vector dimensions (which is even more favorable when similarity calculations are performed in real-time). Finally, by combining the best performing BLL approach with Collaborative Filtering in a round robin fashion, we can recommend jobs with nearly the target novelty $N_{A}$, while still providing high accuracy and diversity.




\section{Conclusion and Future Work}
In this work, we explored the impact of item embeddings for tackling the problem of recommending jobs to university students. We proposed how to integrate the frequency and recency of job interactions when combining embeddings for content-based recommendation. Our evaluation conducted on a dataset gathered from the Austrian student job portal Studo showed that a recommendation approach that uses the BLL equation from human memory theory provides a good balance between accuracy and diversity while achieving the best performance with respect to our adapted novelty measure. Moreover, combining the BLL approach with Collaborative Filtering can lead to a robust model with respect
accuracy, diversity and novelty. 

With respect to our next steps, we plan to perform an online study to asses the real user acceptance of utilizing the BLL equation for constructing item embeddings. Furthermore, we plan to investigate how to integrate university related metadata (e.g., recently visited lectures, expected date of graduation, etc.) within the trained item embeddings in order to better assess what job postings are most relevant for a given student. 

\balance

\bibliographystyle{abbrv}

\begin{thebibliography}{10}

\bibitem{al2012survey}
S.~T. Al-Otaibi and M.~Ykhlef.
\newblock A survey of job recommender systems.
\newblock {\em International Journal of Physical Sciences}, 7(29):5127--5142,
  2012.

\bibitem{anderson2004integrated}
J.~R. Anderson, D.~Bothell, M.~D. Byrne, S.~Douglass, C.~Lebiere, and Y.~Qin.
\newblock An integrated theory of the mind.
\newblock {\em Psychological review}, 111(4):1036, 2004.

\bibitem{huang2016exploiting}
Y.~Huang.
\newblock Exploiting embedding in content-based recommender systems.
\newblock 2016.

\bibitem{jones2014college}
J.~Jones, J.~Schmitt, et~al.
\newblock A college degree is no guarantee.
\newblock Technical report, Center for Economic and Policy Research (CEPR),
  2014.

\bibitem{Kenthapadi2017}
K.~Kenthapadi, B.~Le, and G.~Venkataraman.
\newblock Personalized job recommendation system at linkedin: Practical
  challenges and lessons learned.
\newblock In {\em Proc. of ACM RecSys'17}.

\bibitem{kowald2017temporal}
D.~Kowald, S.~C. Pujari, and E.~Lex.
\newblock Temporal effects on hashtag reuse in twitter: A cognitive-inspired
  hashtag recommendation approach.
\newblock In {\em Proc. of WWW'17}.

\bibitem{lacic2017tailoring}
E.~Lacic, D.~Kowald, and E.~Lex.
\newblock Tailoring recommendations for a multi-domain environment.
\newblock In {\em Proc. of RecSysKTL'17 co-located with ACM RecSys'17}.

\bibitem{le2014distributed}
Q.~Le and T.~Mikolov.
\newblock Distributed representations of sentences and documents.
\newblock In {\em Proc. of ICML'14}.

\bibitem{Liu2016}
K.~Liu, X.~Shi, A.~Kumar, L.~Zhu, and P.~Natarajan.
\newblock Temporal learning and sequence modeling for a job recommender system.
\newblock In {\em Proc. of RecSys Challenge'16}.

\bibitem{Liu2017}
R.~Liu, W.~Rong, Y.~Ouyang, and Z.~Xiong.
\newblock A hierarchical similarity based job recommendation service framework
  for university students.
\newblock {\em Frontiers of Computer Science}, 11(5):912--922, Oct 2017.

\bibitem{mikolov2013distributed}
T.~Mikolov, I.~Sutskever, K.~Chen, G.~S. Corrado, and J.~Dean.
\newblock Distributed representations of words and phrases and their
  compositionality.
\newblock In {\em Advances in neural information processing systems}, pages
  3111--3119, 2013.

\bibitem{musto2016learning}
C.~Musto, G.~Semeraro, M.~de~Gemmis, and P.~Lops.
\newblock Learning word embeddings from wikipedia for content-based recommender
  systems.
\newblock In {\em Proc. of ECIR'16}.

\bibitem{Paparrizos2011}
I.~Paparrizos, B.~B. Cambazoglu, and A.~Gionis.
\newblock Machine learned job recommendation.
\newblock In {\em Proc. of ACM RecSys'11}, New York, NY, USA.

\bibitem{ParraSahebi}
D.~Parra and S.~Sahebi.
\newblock Recommender systems : Sources of knowledge and evaluation metrics.
\newblock In {\em Advanced Techniques in Web Intelligence-2: Web User Browsing
  Behaviour and Preference Analysis}, pages 149--175. Springer, 2013.

\bibitem{pu2011user}
P.~Pu, L.~Chen, and R.~Hu.
\newblock A user-centric evaluation framework for recommender systems.
\newblock In {\em Proc. of ACM RecSys'11}.

\bibitem{SmythMcClave01}
B.~Smyth and P.~McClave.
\newblock Similarity vs. diversity.
\newblock In {\em Proc. of ICCBR '01}.

\bibitem{yuan2016solving}
J.~Yuan, W.~Shalaby, M.~Korayem, D.~Lin, K.~AlJadda, and J.~Luo.
\newblock Solving cold-start problem in large-scale recommendation engines: A
  deep learning approach.
\newblock In {\em Proc. of Big Data'16}. IEEE.

\bibitem{zhang2016ensemble}
C.~Zhang and X.~Cheng.
\newblock An ensemble method for job recommender systems.
\newblock In {\em Proceedings of the Recommender Systems Challenge}, page~2.
  ACM, 2016.

\bibitem{zhou2010solving}
T.~Zhou, Z.~Kuscsik, J.-G. Liu, M.~Medo, J.~R. Wakeling, and Y.-C. Zhang.
\newblock Solving the apparent diversity-accuracy dilemma of recommender
  systems.
\newblock {\em Proceedings of the National Academy of Sciences},
  107(10):4511--4515, 2010.

\end{thebibliography}

\end{document}